\documentclass[numbered]{trbunofficial}
\usepackage{graphicx}
\usepackage{makecell}
\usepackage{amsfonts}
\usepackage{multirow}
\usepackage{amsmath}
\usepackage{relsize}
\usepackage{url}
\usepackage[symbol]{footmisc}

\newcommand{\HD}{\mathsf{HD}}

\usepackage[ruled,vlined,linesnumbered]{algorithm2e}
\usepackage{caption}
\usepackage{subcaption}

\usepackage{comment}

\usepackage{comment}

\usepackage{booktabs,array}

\newcount\rowc

\makeatletter
\def\ttabular{%
\hbox\bgroup
\let\\\cr
\def\rulea{\ifnum\rowc=\@ne \hrule height 1.3pt \fi}
\def\ruleb{
\ifnum\rowc=1\hrule height 1.3pt \else
\ifnum\rowc=6\hrule height \heavyrulewidth 
   \else \hrule height \lightrulewidth\fi\fi}
\valign\bgroup
\global\rowc\@ne
\rulea
\hbox to 10em{\strut \hfill##\hfill}%
\ruleb
&&%
\global\advance\rowc\@ne
\hbox to 10em{\strut\hfill##\hfill}%
\ruleb
\cr}
\def\endttabular{%
\crcr\egroup\egroup}

\usepackage[colorinlistoftodos,prependcaption,textsize=small]{todonotes}

\def\D{{\cal D}}

\def\F{{\cal F}}
\def\G{{\cal G}}

\def\S{{\cal S}}

\def\reals{\mathbb{R}}

\usepackage[hidelinks]{hyperref}

\AuthorHeaders{M. Mirzanezhad, R. Twumasi-Boakye, A. Broaddus, and T. Fabusuyi}
\title{Estimating Demand for Online Delivery using Limited Historical Observations}

\author{
\textbf{Majid Mirzanezhad
} \\
Transportation Research Institute \\
University of Michigan \\
Ann Arbor, MI. 48109, USA.\\
miirza@umich.edu \\
\hfill\break
\textbf{Richard Twumasi-Boakye
}\\
Research Scientist \\
Robotics and Mobility Research \\
Ford Motor Company \\
rtwumasi@ford.com \\
\hfill\break
\textbf{Andrea Broaddus
} \\
Senior Research Scientist \\
Robotics and Mobility Research \\
Ford Motor Company \\
abroaddu@ford.com \\
\hfill\break
\textbf{Tayo Fabusuyi
} \\
Transportation Research Institute \\
University of Michigan \\
Ann Arbor, MI. 48109, USA.\\
fabusuyi@umich.edu
}

\begin{document}
\maketitle

\section{Abstract}

Driven in part by the COVID-19 pandemic, the pace of online purchases for at-home delivery has accelerated significantly. However, responding to this development has been challenging given the lack of public data
. The existing data may be infrequent, and a significant portion of data may be missing because of survey participant non-responses. This data paucity renders conventional predictive models unreliable. 
We address this shortcoming by developing algorithms for data imputation and synthetic demand estimation for future years without the actual ground truth data. We use 2017 Puget Sound Regional Council (PSRC) and National Household Travel Survey (NHTS) data and impute from the NHTS for the Seattle-Tacoma-Bellevue MSA where delivery data is relatively more frequent. Our imputation has the mean-squared error $\mathsf{MSE} \approx 0.65$ to NHTS with mean $\approx 1$ and standard deviation $\approx 3.5$
and provides a similarity matching between the two data sources' samples. 
Given the unavailability of NHTS data for 2021, we use the temporal fidelity of PSRC data sources (2017 and 2021) to project the resolution onto the NHTS providing a synthetic estimate of NHTS deliveries. Beyond the improved reliability of the estimates, we report explanatory variables that were relevant in determining the volume of deliveries. This work furthers existing methods in demand estimation for goods deliveries by maximizing available sparse data to generate reasonable estimates that could facilitate policy decisions.

\hfill\break%
\noindent\textit{Keywords}: Delivery estimation, Travel survey data, Imputation
\newpage

\section{Introduction} \label{sec:intro}

%
%
%
%

In recent years, and especially since the onset of the COVID-19 pandemic, online sales platforms and apps for purchasing goods online have become ubiquitous since they facilitate daily tasks among people in an efficient and effective manner. One of the primary transactions that significantly increased during the pandemic was online package delivery orders. Delivering the online orders in an efficient manner to unique households can be viewed as a crucial computational problem that involves different disciplines. 

Due to the COVID-19 pandemic, the volume of online transactions increased dramatically from 7\% of households conducting online transactions in 2019 to 47\% in 2021~\cite{psrc}. This is particularly noticeable with the online demand for product deliveries such as groceries, store packages and restaurant deliveries. Businesses offering home delivery services, including Amazon Prime, UberEats, DoorDash, Grubhub and others, are interested in predicting online orders ahead of time so that they can better manage inventory and deliveries. While online delivery demand presents business opportunities, we lack publicly available delivery data for predicting and forecasting how many orders might be generated by households in a particular area, thus making it difficult to prototype delivery services for cities. Currently, there are no existing methodologies for solving this problem, in the public realm. Our approach addresses this shortcoming by leveraging predictive variables from multiple datasets to estimate on-demand delivery volumes in a manner that could be easily replicated.
We use data derived from multiple sources to  estimate the number of online purchases for at-home delivery per household, spanning a range of delivery goods such as food, package, and groceries.  Such data can be obtained either implicitly, i.e., through platforms, APIs, apps, or explicitly through questionnaires, surveys, etc. Both methods of collecting the data shed light on the frequency and type of each goods delivery to households.


While online delivery demand presents business opportunities, we lack publicly available delivery data, thus making it difficult to prototype delivery services for cities. Currently, there are no existing methodologies for solving this problem. Our approach addresses this shortcoming by leveraging predictive variables from multiple datasets to estimate on-demand delivery volumes in a manner that could be easily replicated.  
In this paper, we use data from the Puget Sound Regional Council (PSRC) Household Travel Survey  (HTS), including  information on online purchases; the individuals making these purchases and the household they are associated with for the Seattle Metropolitan Statistical area (MSA). This data, augmented with additional data from the 2017 National Household Travel Survey (NHTS)~\cite{nhts}, provides information on the volume of online purchases per household and explanatory variables. The explanatory variables are classified broadly into three types - those related to the online purchase, such as the nature of the product; those related to the individual making the purchase, such as their gender and employment status, and those related to the household that the individual is associated with. 

This paper is organized as follows: Section~\ref{sec:lit-review} discusses related research in online at-home deliveries; Section~\ref{sec:data} examines the data specification exploited in our analysis for this research and discusses the requisite treatment prior to modeling; Section~\ref{sec:method} highlights the  methodological contributions of this study given the current landscape of computational approaches for creating synthetic demand; in particular, Section~\ref{sec:generate} provides an in-depth description of the methodology and algorithms that synthetizes the National Household data given the method presented in the previous section; Section~\ref{sec: experiment} provide some experimental results on our datasets obtained from our methods; Section~\ref{sec:exploratory} details the exploratory results and contributing parameters to the prediction; and Section~\ref{sec:conclusion} concludes the paper with the main research findings and potential future work.

\section{Related work} \label{sec:lit-review}

A decent body of research has been devoted to online at-home deliveries per household and the covariates that determine the volume of the online transactions. Using an online survey collected in June 2020, Dsouza and Sharma~\cite{dsouza-sharma} studied consumer behavior in the context of online food delivery services during the COVID-19 pandemic. The responses collected were analyzed based on the constructs formed for the following tests: reliability, convergent and discriminant analysis. Also, principal component analysis was performed to ensure that the variables are correlated to each other. They indicated that food quality plays a crucial role for customer satisfaction which indirectly influences customer loyalty. Following that, Mehrolia et al.~\cite{mehrolia} empirically measured the characteristics of customers who did  and  did  not  order  food  through  online  food  delivery  services.

Kim and Wang~\cite{KIM2021103052} examined factors affecting the deliveries of online orders in New York city during the COVID-19 pandemic while focusing more on food, grocery, and retail deliveries. Using a seemingly unrelated regression modeling approach, they showed that black males who have children and use smartphones in the New York City borough of Manhattan are more apt to use food delivery apps. On the other hand, being white, with children and having access to a vehicle are the most significant parameters in grocery and retail online deliveries. Unnikrishnan and Figliozzi~\cite{unner-fig-20} studied online survey responses by participants from Vancouver-Hillsboro and Oregon-Washington metropolitan area during COVID-19. They used ordinal logit regression and examined the number of deliveries before and during the pandemic. In their experiment, the number of deliveries increased from approximately 6-10 before pandemic to more than 10 during the pandemic. Income, technology accessibility,and the presence of brick and mortar stores were all reported to be statistically significant variables in predicting online purchases per household. Punel et al.~\cite{PUNEL201830} studied how and to what extent attitudes, preferences, and characteristics of crowd-shipping users differ from non-users, and they concluded that crowd-shipping is more prevalent among young people, men, and full-time employed individuals. In Dhaka city, India, the unreliability of online transactions via credit cards due to scams leads customers to pay cash at the  door for home deliveries~\cite{Anisur2018}. A similar investigation has been conducted in~\cite{Shekhar-18} that shows the usefulness and risk of a product were the top two significant predictors of online purchase intention for Indian consumers.

Nguyen et al.~\cite{Nguyen-19} used a survey-based statistical technique that helps determine how customers value different attributes of a product in terms of feature, function, benefits. They classified these attributes into three categories: price-oriented, time and convenience-oriented, and value-for-money-oriented and found that the most important factor is the delivery fee. Lisnawati et al.~\cite{Lisnawati2020} administered questionnaires to 200 respondents who use online applications for food delivery service such as Grab, McDonald, KFC and Pizza Hut to help identify consumer behavior regarding the use of online food delivery in order to design more effective marketing strategies. 
They observed that young adults, 18-24 years old, making between \$50,000 - \$100,000 income per annum are the most consuming segment of the population for online deliveries. sociodemographic characteristics and the adoption of online grocery shopping were also examined in ~\cite{VanDroogen-17}. The authors showed that the presence of young children and the working situation in the household matter more than the age of the individual. 

A recent work by Fabusuyi et al. estimated demand for online package delivery for households within a small geographic area used a combination of NHTS data and synthetic data obtained from the Southeast Florida Regional Planning Model (SERPM) to generate estimates of online delivery purchases for Miami Dade county at the level of microanalysis zones (MAZs)~\cite{FABUSUYI-2020}. The estimates were validated using measures of predictive accuracy and by evaluating the predicted aggregated values relative to the population estimates generated from the NHTS survey data. 


It is evident from the existing literature that studies on online purchase and at-home deliveries have largely focused on contributory factors to consumers’ choices, such as smartphone availability or the attribute of purchased goods. Other studies identify the socio-economic characteristics of households that lean towards purchasing goods more frequently. The forecasting problem has not been thoroughly invested in existing literature, except in studies by Fabusuyi et al.~\cite{FABUSUYI-2020} who utilized synthetic population data to generate online delivery volumes at the micro analysis zone level for Miami-Dade county, and another recent study by Crivellari et al.~\cite{Crivellari} who leveraged a deep neural-network approach for estimating short term food delivery in urban areas – they used a Convolutional Neural Network - Long Short-Term Memory Networks (CNN-LSTM) regressor trained on time series data to arrive at estimate that could be used to operationalize short deliveries. However, we have a clear research gap with respect to predicting the number of online deliveries in a given area, or the number of online orders made by a household in a given day, especially where we have incomplete data sets. This is the research gap addressed in this paper.

\subsection{Our contribution}
This research moves the needle in demand estimation methods for goods deliveries by presenting novel approaches in two main ways: (i) leveraging robust imputation methods for treating missing observations in public data sets; (ii) algorithms for generating synthetic demand for future years. 
There are few publicly available datasets that include home delivery information. We identified two household travel surveys that asked households to report the number of home deliveries they received, per day (e.g. PSRC and NHTS). This data was suitable for use as the dependent variable, but neither dataset was sufficient, on its own, to build a robust predictive model. 

The temporal resolution of the NHTS data is relatively low - the most recent with the full data complement is from 2017.  However, compared to the 2017 PSRC data, it suffers less from missing data on on-demand deliveries. 
In particular, we propose a method that imputes missing values in the PSRC 2017 data by matching the data to the Seattle area 2017 NHTS dataset. The matching process captures the similarity between the two data samples and project the frequencies from the NHTS samples onto their perfect PSRC matches. We use \emph{Nearest-Neighbor Matching}~\cite{lux-interpolate-highdimension-21, coverhart-1967} between the samples to address the imputation.
Broadly speaking, there are a few drawbacks with conventional imputation methods that use the known values within the dataset: 
(1) Substituting a majority of missing values with average, minimum, maximum, most frequent values does not lead to realistic estimation, since the imputed values typically defaults to the most frequent data points (in our case 0 deliveries); 
(2) Other statistical approaches like the {\em Multivariate Imputation by Chained Equation (MICE)} algorithm by Rubin~\cite{rubin-impute-1987} make a fairly strong assumption on survey data that the values are “missing at random” which may not  be the case, and in the present study, missing values are the variable of interest (delivery), which is the basis of our estimation;
(3) The MICE algorithm uses samples within the same dataset for imputing the missing values and cannot reflect the actual values from another dataset;
(4) these conventional imputation methods may yield negative values, which does not make sense for predicting the number of household deliveries.  

We propose a predictive algorithm suitable for forecasting demand for household deliveries at the zip code level.  Our imputation with high veracity estimation of deliveries in PSRC 2017 has the mean-squared error $\mathsf{MSE} \approx 0.65$ to the ground-truth dataset with mean $\approx 1$ and standard deviation $\approx 3.5$, and the distribution of the deliveries is consistent with the one in NHTS 2017.

Next, we use the imputed matches obtained between the samples of PSRC and NHTS in 2017 and we generate a synthetic dataset of NHTS 2021, given the query dataset PSRC 2021. Then the matching procedure is applied again, match households from the synthetic NHTS 2021 and PSRC 2021 datasets. For any sample in PSRC 2021, we use the similarity measures of the matching and generate a prospective sample that matches the source sample best. We systematically provide experimental evaluations between any pairs of the datasets of different time-segments. Through this process, explanatory variables that were relevant in determining the volume of online purchases such as age, income, education, number of adults in a household were also systematically obtained.


\section{Data} \label{sec:data}
\begin{figure}[htbp]
		\begin{center} 
 			\includegraphics[width = 6.5cm]{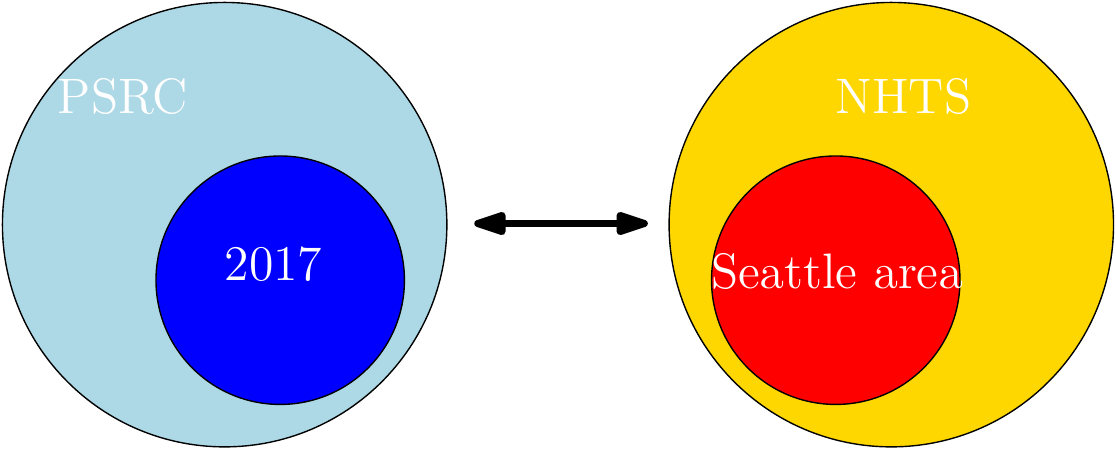}
 			\caption{Panoramic representation of the two datasets. }
			\label{fig:append}
		\end{center}
\end{figure}

As mentioned earlier, we use the Puget Sound Regional Council Travel Survey 2017 (PSRC)~\cite{psrc}. In PSRC 2017 the survey reports responses from 2665 households in the Seattle-Tacoma-Bellevue metropolitan statistical area. Each household provides information about their annual income, individual's employment status, level of education, their online orders, etc. Another primary data source that we use is the National Household Travel Survey 2017 (NHTS)~\cite{nhts}. We intend to exploit those features of the two data sets that are associated with a specific region and time-segment. 
We restrict our analysis to those samples of NHTS in Seattle area of Zip Code ``42660'' (NHTS-cut) which both PSRC and NHTS datasets refer to. Although these two sets of observations may not demonstrate identical households, they allow us to have two pools of samples picked from the same population. Figure~\ref{fig:append} shows the schematic representation of a temporally-restricted chunk of PSRC data and a spatially-restricted chunk of NHTS data.   

The obtained dataset inherits only those features in the two datasets that are {\em conceptually identical}. For the sake of brevity, below we list a subset of the conceptually identical features between the two datasets that appear in PSRC and NHTS both:

$$ {Features} = \{ \mathsf{Income,~Age,~Sex,~Education,~no.~Adults/no.~Children,~Employment,~Delivery} \}.$$

\begin{figure}[htbp]
		\begin{center} 
 			\includegraphics[width = 8.5cm]{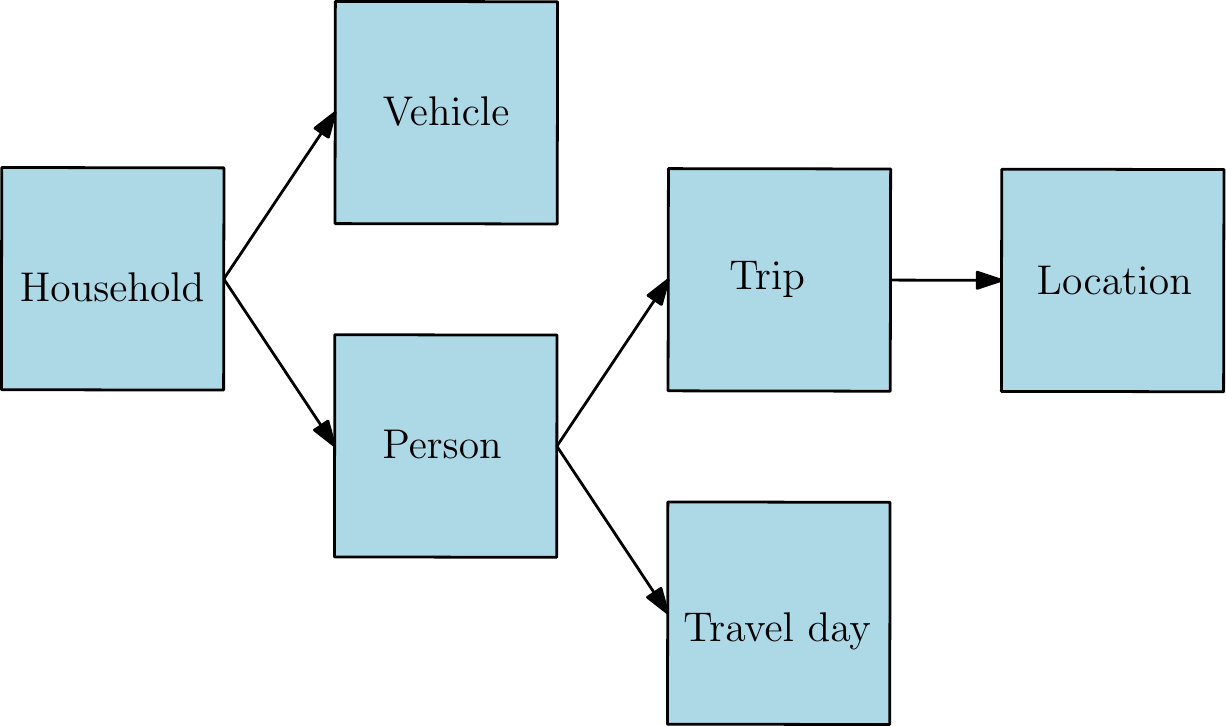}
 			\caption{The hierarchy of data tables in PSRC and NHTS travel surveys.}
			\label{fig:psrc_hierarchy}
		\end{center}
\end{figure}

Note that every household consists of multiple individuals living in the house, and each person has reported their own information in the survey see Figure~\ref{fig:psrc_hierarchy}). 
Therefore each household consists of multiple samples in our datasets. The \emph{Delivery} feature is presented in Travel Day table and the actual number of deliveries per household is the aggregation of the amount of deliveries over all samples in Travel Day and Person tables associated with the household.  

Each of our datasets is a unified dataset comprised by joining all tables presented in Figure~\ref{fig:psrc_hierarchy} linked through their primary keys (IDs). Some statistics on the size of the datasets and their missing percentile are described in Table~\ref{tab:desc_stats}.  Note that we refer to the Seattle-area cut of the NHTS dataset used in this analysis as “NHTS-cut”.
\begin{table}[!h]
	\centering 
	\begin{tabular}[c]{|c|c|c|c|c|}
		\hline 
Dataset &   Households & Samples &  \makecell{Samples with \\ missing deliveries} & \makecell{Percentile of the \\ missing deliveries} \\ \hline
PSRC 2017 & 2665 & 363970 & 349787 & 96\%   \\ \hline
NHTS-cut 2017  &  272 & 4063 & 322 & 7\% 
\\ \hline

PSRC 2021 & 2020 & 20797 & 8780 & 29\%   \\ \hline

	\end{tabular}
		
	\caption{Statistics on the size of our data sources. Missing percentile is obtained by the number of missing delivery volumes divided by the number of samples.}
	\label{tab:desc_stats}
\end{table}

We observe that most of the delivery data (our dependent variable) in PSRC 2017 are missing and this scarcity substantially appears to be much less in PSRC 2021. More detailed statistics of the datasets are provided in Table~\ref{tab:desc_stats2}.

\begin{table}

	\centering 
\begin{tabular}{|c|c|c|c|}
\hline
Data Source & PSRC 2017 & NHTS-cut 2017 & PSRC 2021 \\\hline
Income: \makecell[c]{
$>$ \$100,000:  \\
\$75,000 - \$100,000:  \\
$<$ \$75,000:  \\ 
Missing:  \\
Total households:  }  & \makecell[c]{
1171 \\ 
369 \\
941 \\
184 \\
2665} & 
\makecell[c]{
88 \\ 
70 \\
106 \\
8 \\
272} & 
\makecell[c]{
340 \\ 
154 \\
557 \\
557 \\
2020}
\\ \hline
Age: \makecell[c]{
$<$ 25:  \\
25 - 45:  \\
45 - 65:  \\
$>$ 65:  \\
Missing:   \\
Total individuals: } & 
\makecell[c]{
940  \\
2282  \\
1122  \\
580  \\
0   \\
4924 } & 
\makecell[c]{
67  \\
161  \\
156  \\
104  \\
0   \\
488 } &
\makecell[c]{
72  \\
414  \\
347  \\
284  \\
2056   \\
3173 } \\
\hline
Education: 
\makecell{
$<$ high school:  \\
High school grad.:  \\
Technical training: \\
Associate degree:   \\
Bachelor degree:    \\
Graduate degree:    \\
Missing:  \\
Total individuals: } &
\makecell{
 40    \\
 203 \\
 98 \\
 644 \\
 1762 \\
1472 \\
 705\\
 4924 }
& \makecell{
17 \\
46  \\
0  \\
99  \\
162  \\
122 \\
42  \\
488  } & 
\makecell{
18 \\
116  \\
38  \\
352  \\
346  \\
247 \\
2056  \\
3173  }
\\
\hline
Gender: \makecell{
Female:  \\
Male:   \\
Missing:   \\
Total individuals:  } &
\makecell{
 2440 \\
2396 \\
88  \\
 4924  }
& \makecell{
257 \\
231 \\
0  \\
488 }
& \makecell{
593 \\
502 \\
2078  \\
3173 }  \\
\hline
Life cycle: \makecell{
2 adults, no children:  \\
1 adult, no children:  \\
1 adult, with children: \\
2 adults, with children: \\
Missing:   \\
Total households: } & 
\makecell{
1254 \\
880 \\
531 \\
0 \\
0 \\
2665 }
& \makecell{
124 \\
73  \\
6  \\
69 \\
0 \\
272 }
&  \makecell{
1 \\
1  \\
1  \\
1113 \\
904 \\
2020 }
\\
\hline
Employment: \makecell{
Full time:  \\
Retired:   \\
Part time:  \\
Freelancer:  \\
Not employed:  \\
Homemaker:  \\
Volunteer(intern):  \\
Missing: \\
Total individuals:  }  & 
\makecell{
2600 \\
538 \\
344 \\
268 \\
239 \\
197 \\
 33\\
 705 \\
 4924  }
& 
\makecell{
236 \\
0 \\
45 \\
0 \\
0 \\
0 \\
0\\
207 \\
488 }
& 
\makecell{
527 \\
254 \\
100 \\
59 \\
99 \\
47 \\
8 \\
2079 \\
3173 } \\
\hline
\end{tabular}
		
	\caption{Descriptive statistics of the features in PSRC and NHTS data sources.}
	\label{tab:desc_stats2}
\end{table}

\section{Methodology}~\label{sec:method}
The main ingredient of our method boils down to matching techniques. Given two {\em pools} of samples, the primary approach is to match the samples of the two pools based on the similarity of their attributes. Once matched, we propagate the value of deliveries from a sample in the groundtruth dataset to the matched ones in the dataset of interest (PSRC 2017). Our ground-truth dataset contains all samples with available ``Delivery'' values. We concatenate the NHTS-cut 2017 and the portion of PSRC 2017 with present ``Delivery'' values and represent the obtained dataset as the ground-truth.    
For brevity, throughout the paper, the ground-truth dataset is said to be {\em candidate} and the one that is supposed to be imputed, is said to be {\em source}.

\subsection{Data Cleansing}
In order to make use of another dataset to impute, we must make the two datasets consistent. This means that we only concentrate on those features that are in {\em conceptually identical} between the two datasets. 
Two features, each from one dataset, are conceptually identical if their corresponding sets of categorical values overlap, but may not necessarily be the same categories. 

\subsubsection{Harmonization}
Since the features of the two datasets may have different sets of categories, we have to unify their categories in order to perform any computations on them, such as matching. This process is called data harmonization. After the harmonization process, the observations per feature across the datasets become consistent in terms of the categories they represent. For instance, the features “delivery\_pkgs\_freq”, “delivery\_food\_freq” and “delivery\_grocery\_freq”  in PSRC and “DELIVER” in NHTS conceptually refer to the number of deliveries. However, the summation of “delivery\_pkgs\_freq”, “delivery\_food\_freq” and “delivery\_grocery\_freq” refers to the total number of deliveries on the travel day reported by the survey respondent while  “DELIVER” counts the total number of deliveries per month. Therefore they are conceptually identical and they appear as one feature namely `Delivery' in our cleansed dataset. In order to harmonize the values in `Delivery' feature, we have to divide each observation with “DELIVER” feature by 30 days to have the average frequency of deliveries per day in order to make it consistent with the features in PSRC. We do this for all common features prior to our estimation.


\subsubsection{Imputation of Categorical Features} \label{subsec:impute-cat}

Categorical data are the most commonly used data types that we deal with in our datasets. Participants to the survey may not select all the checkboxes in the PSRC and NHTS questionnaire, thus this causes missing (NaN) values in our records.  Categorical features consist of non-numeric data types such as strings, data/time, and others and statistical and machine learning models are not capable of working with such data types.
In this case, since there are presumed values per sample, we binarize the feature using different {\em bins} and concatenate the name of each value to the end of the feature’s tag. This way, if for a sample, category $c$ of the feature $A$ is selected, we fill the entry of the feature $A_c$ with 1, and 0 otherwise. The high-level structure of a categorical data set is described below:

\begin{table}[h]
    \begin{subtable}[h]{0.5\textwidth}
	\centering 
	\begin{tabular}[c]{|c|c|}
		\hline 
		Features & Categories  \\ \hline \hline
		$A$ & $c, d$ \\ \hline
		$B$ & $i, j$ \\ \hline
		
	\end{tabular}
       \caption{}
       \label{tab:week1}
    \end{subtable}
    \hfill
    \begin{subtable}[h]{0.5\textwidth}
 	\centering 
	\begin{tabular}[c]{|c|c|}
		\hline 
		Alternative features & New categories  \\ \hline \hline
		$A_c$& $0, 1$ \\ \hline
		$A_d$ & $0, 1$ \\ \hline
		$B_i$ & $0, 1$ \\ \hline
		$B_j$ & $0, 1$ \\ \hline
		
	\end{tabular}
        \caption{}
        \label{tab:week2}
     \end{subtable}
     \label{tab:binarize}
\end{table}

For instance, for a sample $s$ the feature $A_c$ is set to $1$ if $s$[$A$] = $c$ and $0$, otherwise. This method can perfectly handle NaN values as well; suppose $s$[$A$] = NaN, our sample after binarization would be $s$[$A_c$] = 0, and $s$[$A_d$] = 0.

This clearly increases the number of features overall, thus leading to having a large size of dataset which cannot be space efficient, however, on the bright side all Nan values will disappear. 
While this method is only useful for categorical data, unfortunately this it may not apply for numerical data type since their affinity can result in tremendously many additional features to the dataset. Therefore, we use a different method to cope with that case.

\subsection{Imputation from a Ground-Truth Dataset}

In this section, we describe our imputation method in more details. We are given two pools of samples collected from the same population within the same time-segment. 
The main idea of our approach is to match the two pools of samples from the same area, based on their similarity and project the  delivery values of the ground-truth sample to the sample of interest. More formally, suppose sample $s$ have a set of features $X$ and the target feature $y$, where $X$ are covariates and $y$ is the delivery value. Thus $s[X]$ indicates all values associated with the feature set $X$ and $s[y]$ refers to the number of delivery values of sample $s$. 

We use the ground-truth (candidate) dataset which is the NHTS-cut 2017 sample and the portion of PSRC 2017 (source dataset) that has present values to impute the missing values of PSRC 2017. Our main method is {\em Nearest Neighbor (NN) Matching} from source dataset (PSRC 2017) to candidate dataset  that captures the similarity between the samples of the two datasets with respect to their corresponding features. The evaluation metric used to assess the closeness of fit for this matching is the Hamming distance between the two feature sets. The nearest neighbor matching gives us a matching that matches each sample in our source dataset to a sample in the candidate dataset. 

Since our features are categorical and are associated with bins, each sample can be considered as a binary vector in $d$ dimension, where $d$ is the number of features.
Suppose we are given two binary vectors  $A= (\{0,1\}^d)$, and $B = (\{0,1\}^d)$ with the same number of coordinates.
The hamming distance $\mathsf{HD}$ between them is defined as:

\[\HD(A,B) := 
\frac{\mbox{number of unequal corresponding coordinates between}~A~\mbox{and}~B}{\mbox{number of coordinates in}~A~\mbox{or}~B}.\]
\\

We present the pseudocode of our algorithm below:


\begin{algorithm}[htbp]
	\DontPrintSemicolon
	\SetKwFunction{Imputation}{\textsc{ImputationAlgorithm}}
	\caption{Imputation Algorithm}
	\label{alg:impute}
	
	\BlankLine 
	\Imputation{$source, candidate, X, y$}:
	
	Group all identical samples together in $candidate[X]$. This gives us a set of buckets $buckets =\{b_1, \cdots, b_k\}$, where $b_j \subseteq candidate$, with $1\leq j\leq k$.

		\ForAll{$bucket \in buckets$}
	{
	$bucket[y] \leftarrow \frac{\sum_{b \in bucket} b[y]}{|bucket|}$
	
	}
	
	Run \textsc{NearestNeighbor} algorithm from $source[X]$ to $buckets[X]$ dataset under the Hamming distance between their samples. This obtains us a nearest neighbor matching $\mu:source \rightarrow buckets$, where $\mu(s_i)=b_j$ under $\HD(s_i,b_j)$ with $1\leq i\leq |source|$ and $1\leq j\leq k$.

	$w \leftarrow \frac{|source|}{|candidate|}$
	
	\ForAll{$sample \in source$}
	{
	$sample[y] \leftarrow {\mu(sample)[y]}/{w}$
	}
	
	\ForAll{$h \in households$}
	{$\sum\limits_{\substack{sample \in h}} sample[y]$}
	
\end{algorithm}

\begin{figure}[!h]
     \centering
     \begin{subfigure}[htpb]{0.55\textwidth}
         \centering
         \includegraphics[width=\textwidth]{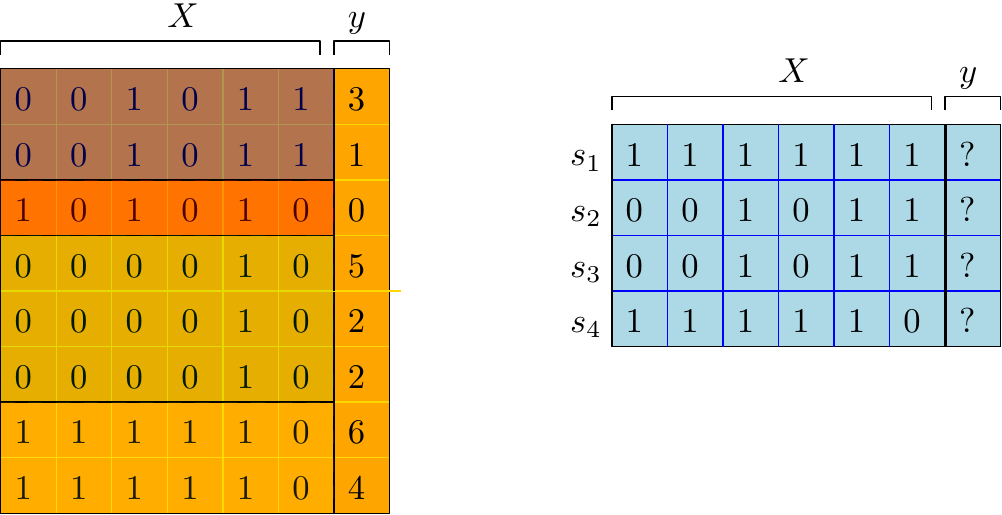}
         \caption{}
         \label{fig:matching1}
     \end{subfigure}
     \hfill
     \begin{subfigure}[htpb]{0.45\textwidth}
         \centering
         \includegraphics[width=\textwidth]{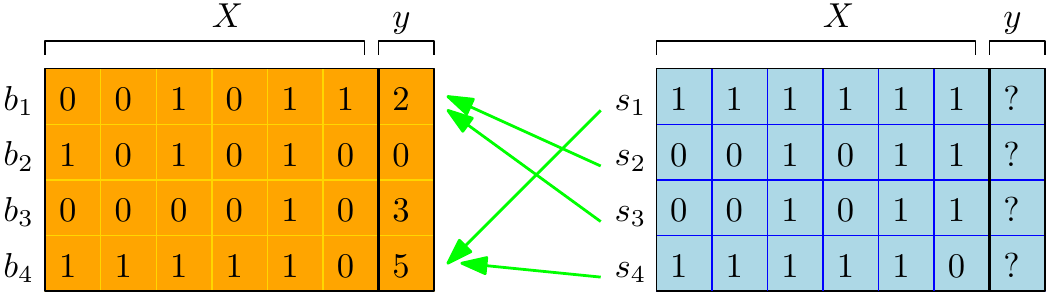}
         \caption{}
         \label{fig:matching2}
     \end{subfigure}
     \hfill
     \begin{subfigure}[htpb]{0.45\textwidth}
         \centering
         \includegraphics[width=\textwidth]{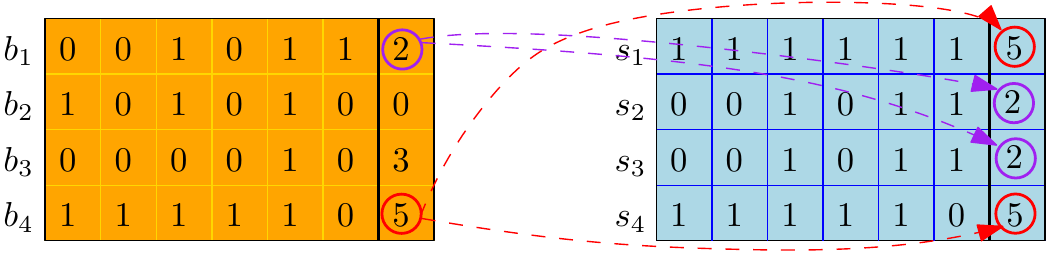}
         \caption{}
         \label{fig:matching3}
     \end{subfigure}
    \caption{(a) We group a set of identical candidate samples by their covariates $X$ and we take the average of the samples' $y$ values. (b) We then perform the Nearest-Neighbor matching from source to the buckets. (c) In the end, we propagate the $y$ values of each bucket to the corresponding matched sample in the source. For simplicity, here we assume that $w = 1$.}
    \label{fig:matching}
\end{figure}

First, we group a set of identical samples of the $candidate$ dataset with respect to their covariate set $X$ and we take the average of their $y$ values. We refer to these as bucket samples, where the number of buckets is the same as the number of groups of different samples. Note that bucket samples will reduce the possibility of a $source$ sample being matched to only one of the identical $candidate$ samples whose $y$ values are different.
We then apply the \textsc{NearestNeighbor} algorithm from $source$ to $candidate$. Since multiple samples from $source$ can be matched to one bucket, we have to appropriately weight $y$ values in $source$, otherwise multiple samples belonging to one household in $source$ inherit the $y$ value of a bucket which inflates the $y$ value for that household.  In a well-behaved dataset (our dataset) the weight $w$ is roughly equal to the size of $source$ divided by the size of $candidate$. Empirically, in our dataset we have that $w\approx 45$.  In the end, we sum up all $y$ values of the samples that belong to the same household.  
An illustration of Algorithm~\ref{alg:impute} is depicted in  Figure~\ref{fig:matching}.

\subsection{Generating National Household Dataset at a Future Time Segment} \label{sec:generate}

In the previous section, we showed that given two datasets collected from the same population and time-frame, we can make use of Nearest-Neighbor (NN) matching and impute one of the datasets whose values are largely missing. In this section, given the source dataset at a future time point in hand (PSRC 2021), we show how to generate the other dataset at the future time point assuming that the same matching in 2017 is preserved between the datasets in 2021.

\begin{figure}[htpb]
    \centering
    \includegraphics[width=0.35\textwidth]{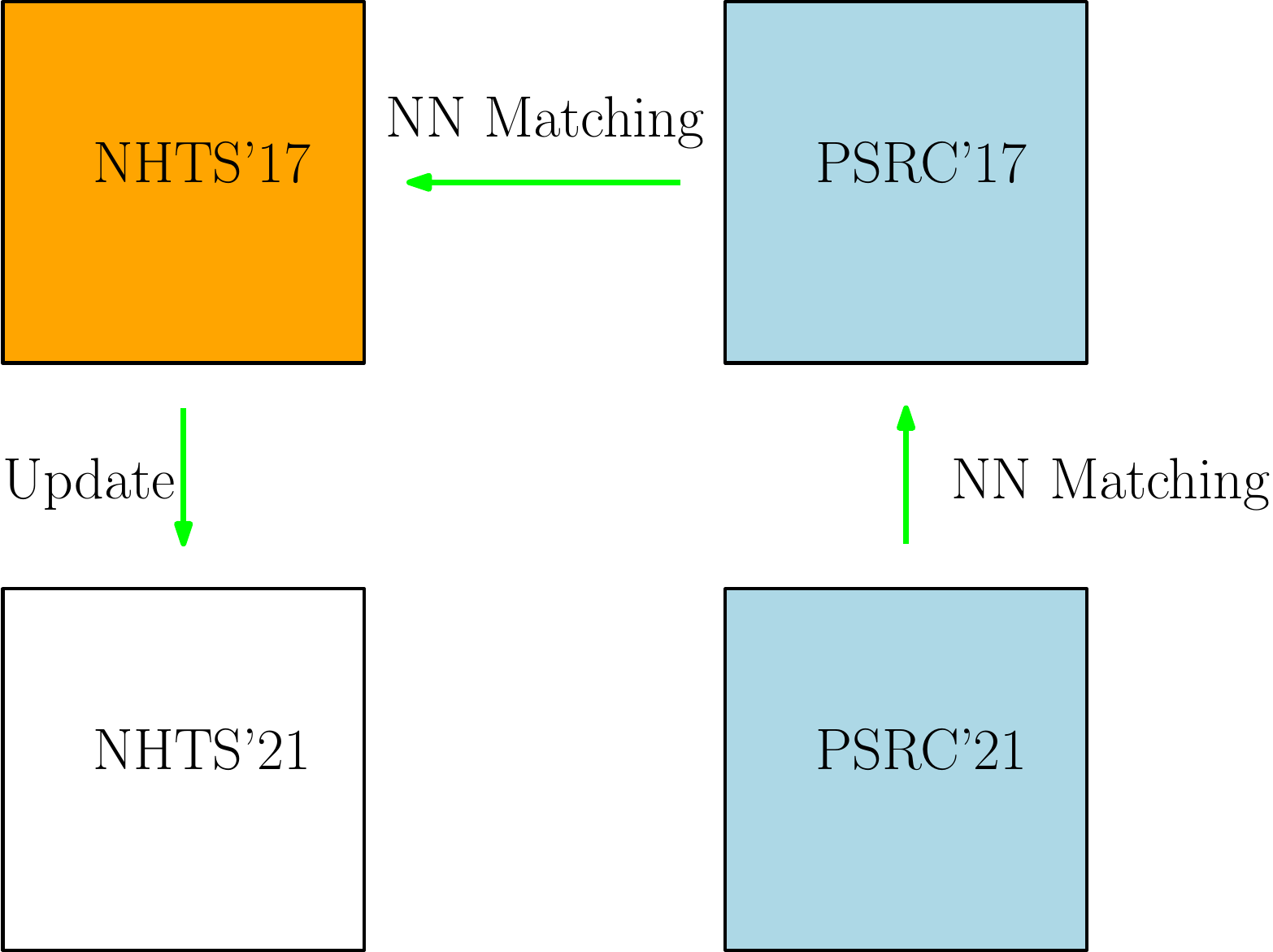}
    \caption{Given PSRC 2021, the core idea is to generate a synthetic  NHTS-cut 2021 based on applying a Nested Nearest Neighbor matching across the datasets and updating the values of NHTS 2017 under the matching obtained.}
    \label{fig:generate_sketch}
\end{figure}

The main idea is to employ a \textsc{Nested Nearest Neighbor} (NNN) matching between the PSRC 2021 and PSRC 2017 and consequently between PSRC 2017 and NHTS-cut  2017 to update the values in NHTS-cut 2017 given the samples in PSRC 2021. For this, we require that the covariates $X$ to be consistent and harmonized across all three datasets (PSRC 2021, PSRC 2017 and the NHTS-cut 2017). Figure~\ref{fig:generate_sketch} shows how to make a bridge between the new dataset and the previous ones in order to generate NHTS 2021.
Algorithm~\ref{alg:generate} presents our idea in pseudocode.

\begin{algorithm}[h]
	\DontPrintSemicolon
	\SetKwFunction{Generate}{\textsc{GenerateAlgorithm}}
	\caption{Generate Algorithm}
	\label{alg:generate}
	
	\BlankLine 
	\Generate{$source_1, source_2, candidate, X, y$}:
	
	Group all identical samples in $candidate[X]$ together. This gives us a set of buckets $buckets =\{b_1, \cdots, b_k\}$, where $b_j \subseteq candidate$ consists of a subsets of samples, with $1\leq j\leq k$. \label{step:grouping}
	
		\ForAll{$bucket \in buckets$}
	{
	$bucket[y] \leftarrow \frac{\sum_{b \in bucket} b[y]}{|bucket|}$ \label{step:bucketmean}
	
	}
	
	Run \textsc{NestedNearestNeighbor} (NNN) algorithm: Apply \textsc{NearestNeighbor} algorithm from $source_2$ to $source_1$ and another \textsc{NearestNeighbor} algorithm from $source_1$ to $buckets$. This gives us two matchings $\mu_2 : source_2 \rightarrow source_1$ and $\mu_1: source_1 \rightarrow buckets$. The nested matching is $\mu := \mu_1 \mathsf{o} \mu_2$. \label{step:NNN}

	$w_1 \leftarrow \frac{|source_1|}{|candidate|}$ and $w_2 \leftarrow \frac{|source_2|}{|source_1|}$ 
	
    $V_1 \leftarrow buckets$, $V_2 \leftarrow source_1$, and $V_3 \leftarrow source_2$
    
    $E_1 \leftarrow \mu_1$, and $E_2 \leftarrow \mu_2$
    
	Build a tri-partite graph $G = (V,E)$, with $V = V_1 \cup V_2\cup V_3$,  and $E = E_1 \cup E_2$.  \label{step:tri-partite}

	$\mathsf{D} \leftarrow \emptyset$
	
	\label{step:loop}
	\ForAll{$v_1 \in V_1$} 
	{
	
	\If{$\exists~v_3 \in V_3$ such that $v_1$ is reachable from $v_3$}
	{

	$\S = \{s~|~\forall s \in V_2~\mbox{s.t.}~ \exists E_1(s, v_1)\}$
	
	\ForAll{$s \in \S$}
	{
		$\G_s = \{g~|~\forall g \in V_3~\mbox{s.t.}~ \exists E_2(g,s)\}$
	}

		$v_1[y] \leftarrow {\mathlarger{\sum}_{\substack{s \in V_2}}\Big(w_1\times \big(\sum\limits_{\substack{g \in \G_s}} (w_2 \times g[y]})/ {|\G_s|}\big)\Big)/{|V_2|}$ \label{step:updateformula}
	
    	$\mathsf{D} \leftarrow  \mathsf{D} \cup v_1$
	
	}

		}
		
return $\mathsf{D}$

\end{algorithm}

\begin{figure}[htpb]
    \centering
    \includegraphics[width=0.85\textwidth]{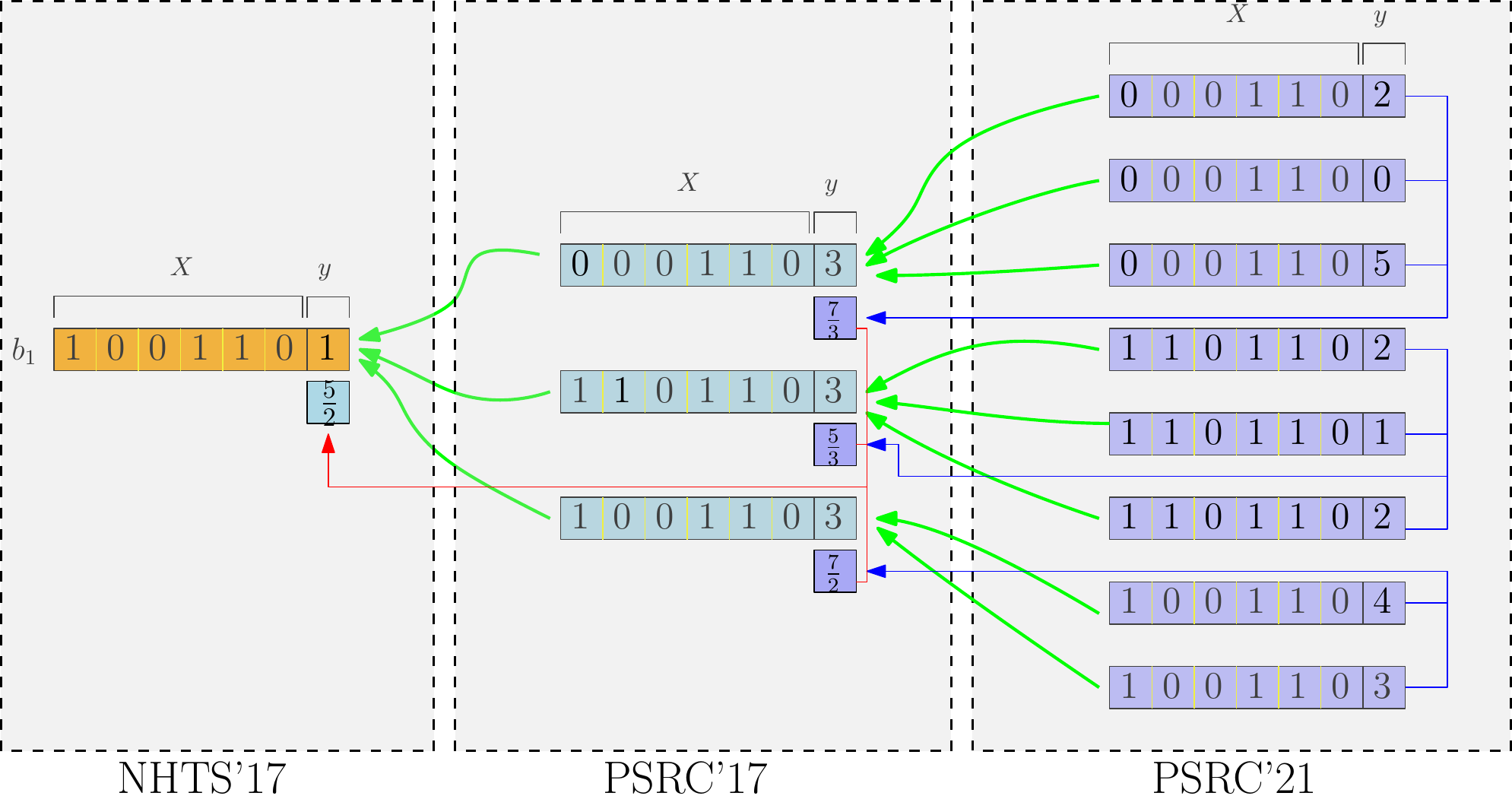}
    \caption{The nested matching between the three datasets forms a tri-partite graph. The average values of the samples are computed in a hierarchical fashion.}
    \label{fig:tri-partite}
\end{figure}

The input data to Algorithms~\ref{alg:generate}, is $source_1$ (PSRC 2017), $source_2$ (PSRC 2021), and $candiates$ (NHTS-cut 2017). The covariate set taken from these datasets is defined as $X$ and the variable of interest is $y$.  
Lines~\ref{step:grouping} - \ref{step:bucketmean} in Algorithm~\ref{alg:generate} are similar to the ones in Algorithm~\ref{alg:impute}. In line~\ref{step:NNN} we perform a nested nearest-neighbor algorithm across all three datasets. This gives us an abstract tri-partite graph $G$ (line~\ref{step:tri-partite}) whose nodes are samples and its edges are the NN matching. We incrementally construct our synthetic dataset $D$. In lines~\ref{step:loop} - \ref{step:updateformula}, for each bucket sample (bucket node) that is connected to at least one sample (one node in $V_3$) in PSRC 2021, we group the nodes in $V$ in a hierarchical manner. We then calculate the weighted average of all samples that are matched to the same node in a higher hierarchical level and we return the values; as shown in Figure~\ref{fig:tri-partite}.

\section{Experimental results} \label{sec: experiment}
In this section we provide some experimental results on our datasets obtained from Algorithm~\ref{alg:impute} and Algorithm~\ref{alg:generate}. We systematically run the datasets through our algorithms and discuss the the effectiveness of each method.   

\subsection{Experiments using Algorithm~\ref{alg:impute}}
\begin{figure} [!h]
     \centering
     \begin{subfigure}[htpb]{0.45\textwidth}
         \centering
         \includegraphics[width=\textwidth]{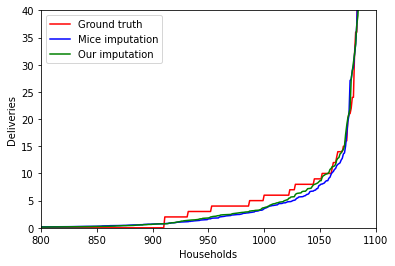}
         \caption{}
         \label{fig:impute_random1}
     \end{subfigure}
     \hfill
     \begin{subfigure}[htpb]{0.45\textwidth}
         \centering
         \includegraphics[width=\textwidth]{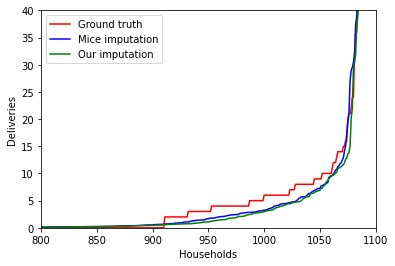}
         \caption{}
         \label{fig:impute_random2}
     \end{subfigure}
     \hfill
     \begin{subfigure}[htpb]{0.45\textwidth}
         \centering
         \includegraphics[width=\textwidth]{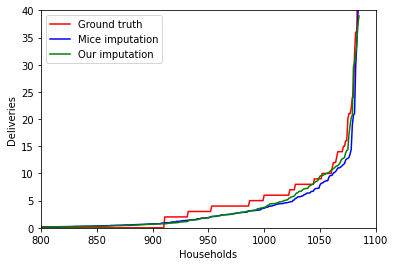}
         \caption{}
         \label{fig:impute_random3}
    \end{subfigure}
    \hfill
     \begin{subfigure}[htpb]{0.45\textwidth}
         \centering
         \includegraphics[width=\textwidth]{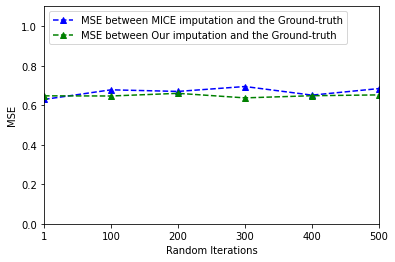}
         \caption{}
         \label{fig:MSE17}
     \end{subfigure}
         \hfill
     \begin{subfigure}[htpb]{0.45\textwidth}
         \centering
         \includegraphics[width=\textwidth]{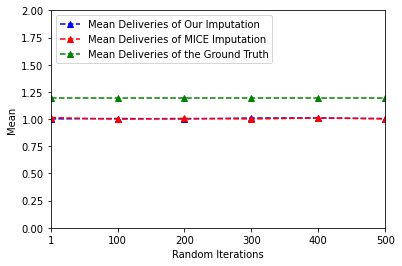}
         \caption{}
         \label{fig:Mean17}
     \end{subfigure} 
         \hfill
     \begin{subfigure}[htpb]{0.45\textwidth}
         \centering
         \includegraphics[width=\textwidth]{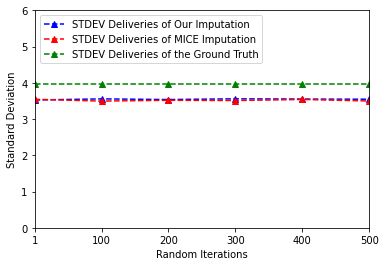}
         \caption{}
         \label{fig:STDV17}
     \end{subfigure}
        \caption{The comparison between our similarity-based imputation, MICE imputation and the ground-truth delivery values over three random subsets of households. Since nearly 800 of the households have delivery of 0, we focus on the interval $[800, 1100]$ where there is more variety of deliveries. In all three cases we observe the distribution of the deliveries are consistent. 
        $\mathsf{MSE}$ between our imputed source dataset and the ground-truth is obtained. These errors are taken over different iteration cutoffs that each is the number of times that we randomly picked subsets of source households and take the average of the MSEs to the gound-truth dataset.}
        \label{fig:impute_compare}
\end{figure}

We aim to show how our imputation algorithm performs through a series of experiments. Since the number of households in our source dataset (PSRC 2017) is  larger (2665 households) than the number of household in the ground-truth dataset (1100 households), we pick a number of random subsets of households in our source dataset whose size is as large as the size of ground-truth dataset and compare the distribution of the delivery ($y$) values. Because there is no unique subset of samples in our source dataset, everytime our picked subset contains different samples. Thus, we devote different cutoffs of the number of time we randomly pick subsets of samples from the source dataset. These cutoffs range from 100 to 500 random iterations. 
We also compare our method with MICE imputation algorithm by~\cite{rubin-impute-1987}. In our experiment, MICE imputation uses {\em Bayesian Ridge} regression, which is a robust regularized linear regression imputation method. The estimation of the model is done by iteratively maximizing the marginal log-likelihood of the available values~\cite{bayesianridge-krushke}. Note that since the households in the two datasets are not identical, we ascendingly sort the delivery values of each and plot them (see Figure~\ref{fig:impute_random1}, ~\ref{fig:impute_random2},~\ref{fig:impute_random3}).

While the plots show a reasonable visual comparison, we also take the average of mean square error ($\mathsf{MSE}$) of the values, as a measure of goodness of fit, over the number of iterations. We do the same for mean and standard deviation of the imputed values (see Figure~\ref{fig:MSE17},~\ref{fig:Mean17},~\ref{fig:STDV17}). As mentioned, since households are not identical, we may not be able to compare the delivery values of corresponding households in the two datasets and take $\mathsf{MSE}$, however, to capture the global behavior of the values, we can sort the values and take the $\mathsf{MSE}$ of the sorted values as well as the mean and standard deviation. For example, in Figure~\ref{fig:MSE17}, the $\mathsf{MSE} \approx 0.65$ and with mean of the delivery values is approximately $1$ and also standard deviation of them is approximately $3.5$ on average over 500 random iterations.

\subsection{Experiments using Algorithm~\ref{alg:generate}}
\begin{figure} [htpb]
     \centering
     \begin{subfigure}[htpb]{0.45\textwidth}
         \centering
         \includegraphics[width=\textwidth]{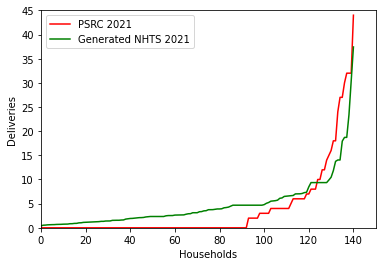}
         \caption{}
         \label{fig:generate_random1}
     \end{subfigure}
     \hfill
     \begin{subfigure}[htpb]{0.45\textwidth}
         \centering
         \includegraphics[width=\textwidth]{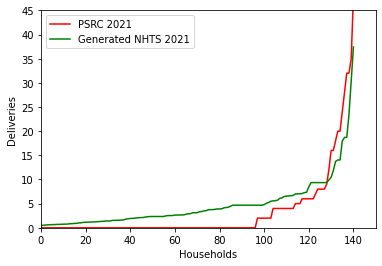}
         \caption{}
         \label{fig:generate_random2}
     \end{subfigure}
     \hfill
     \begin{subfigure}[htpb]{0.45\textwidth}
         \centering
         \includegraphics[width=\textwidth]{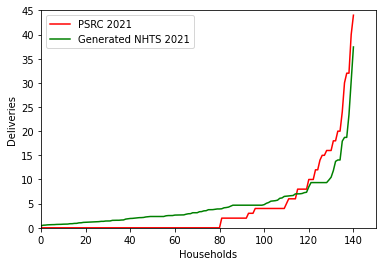}
         \caption{}
         \label{fig:generate_random3}
    \end{subfigure}
    \hfill
     \begin{subfigure}[htpb]{0.45\textwidth}
         \centering
         \includegraphics[width=\textwidth]{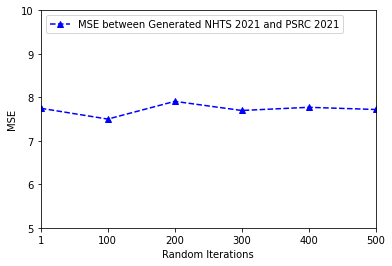}
         \caption{}
         \label{fig:MSE21}
     \end{subfigure}
        \caption{The distributions of the deliveries of three random subsets of 141 households in PSRC 2021 are plotted in (a), (b), and (c) for 141 households. The $\mathsf{MSE}$ in 2021 becomes roughly 10 times larger than the one in 2017.}
        \label{fig:generate_compare}
\end{figure}

\begin{figure} [htpb]
     \centering
     \begin{subfigure}[htpb]{0.45\textwidth}
         \centering
         \includegraphics[width=\textwidth]{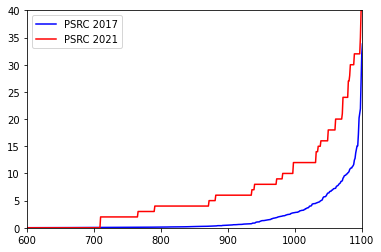}
         \caption{$\mathsf{MSE}= 19.21$}
         \label{fig:generate_random1}
     \end{subfigure}
     \hfill
     \begin{subfigure}[htpb]{0.45\textwidth}
         \centering
         \includegraphics[width=\textwidth]{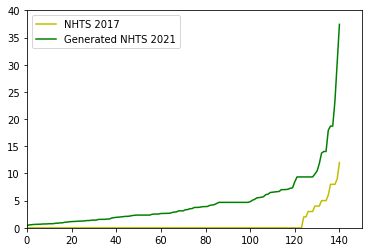}
         \caption{$\mathsf{MSE}= 29.99$}
         \label{fig:generate_random2}
     \end{subfigure}
     \hfill
     \begin{subfigure}[htpb]{0.45\textwidth}
         \centering
         \includegraphics[width=\textwidth]{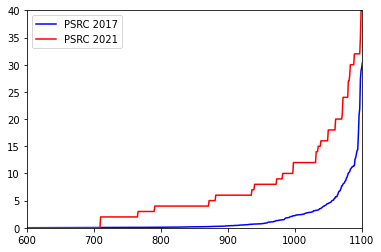}
         \caption{$\mathsf{MSE}= 21.96$}
         \label{fig:generate_random3}
    \end{subfigure}
    \hfill
     \begin{subfigure}[htpb]{0.45\textwidth}
         \centering
         \includegraphics[width=\textwidth]{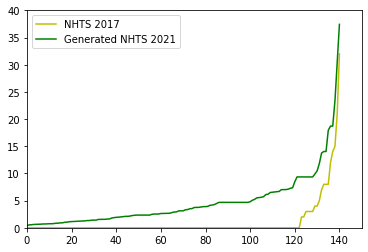}
         \caption{$\mathsf{MSE}= 19.05$}
         \label{fig:MSE21}
     \end{subfigure}
          \begin{subfigure}[htpb]{0.45\textwidth}
         \centering
         \includegraphics[width=\textwidth]{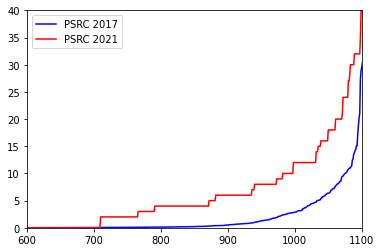}
         \caption{$\mathsf{MSE}= 18.41$}
         \label{fig:generate_random3}
    \end{subfigure}
    \hfill
     \begin{subfigure}[htpb]{0.45\textwidth}
         \centering
         \includegraphics[width=\textwidth]{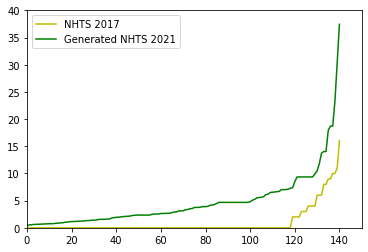}
         \caption{$\mathsf{MSE}= 24.66$}
         \label{fig:MSE21}
     \end{subfigure}
        \caption{There are 3 random subsets of households  in the PSRC 2021 and PSRC 2017 datasets that are compared to the synthetic NHTS-cut 2021 and NHTS-cut 2017 in terms of the spikes (delivery volume increase) on the number of deliveries. The spikes between NHTS-cut datasets during 2017-2021 is consistent with the PSRC datasets. Spikes are measured by $\mathsf{MSE}$ between ascending sorted delivery values.}
        \label{fig:spikes}
\end{figure}

Our experimental analysis of Algorithm~\ref{alg:generate} breaks down into two main parts: (1) the comparison between the PSRC 2021 and the generated synthetic NHTS-cut 2021 (results are shown in Figure~\ref{fig:generate_compare}); and (2) spike analysis of these results (shown in Figure~\ref{fig:spikes}). The spike refers to the significant increase in delivery values between PSRC 2017 and PSRC 2021 due to the temporal gap, which is consistent with the spike between NHTS-cut 2017 and generated synthetic NHTS-cut 2021. Similar to previous section, three random subsets of households in PSRC 2021 are selected and compared to the generated NHTS-cut 2021 (Figure~\ref{fig:generate_random1}, \ref{fig:generate_random2}, \ref{fig:generate_random3}). However, the $\mathsf{MSE}$ between the datasets NHTS-cut 2021 and PSRC 2021 becomes almost 10 times larger than the one between the 2017 datasets. 
From these results, we conclude that Algorithm~\ref{alg:generate} successfully synthesized the data of 141 households from 272 households in NHTS-cut 2017, thus almost more than half of the households were generated in 2021. 
For the sake of evaluation, because the actual NHTS 2021 is not available to compare our synthetic dataset with, we use the spikes between the datasets of the same travel survey to capture the global behavior of delivery values. The idea is to realize if these spikes between each two datasets of the same travel survey are increasing.  We can observe that the spikes between random subsets of households of NHTS-cut 2017 and the synthetic NHTS-cut 2021 is increasing as well as the spike between PSRC 2017 and PSRC 2021. These spikes are measure by $\mathsf{MSE}$ (see Figure~\ref{fig:spikes}).

\newpage
\section{Affecting parameters on delivery values} \label{sec:exploratory}

Finally, we provide an exploratory analysis on our datasets each to reveal covariates that contribute to the estimation the most. 
We highlight some high-level information obtained from our model. In order to achieve this, we need to train our imputed values by a machine learning model since it provides us with a function that describes each feature's contribution to the imputed delivery values. We use the gradient descent boosting regression model, XGBoost in Python 3.10~\cite{xgboost-Chen-Guestin}. We train our datasets exclusively and compute a regression that minimizes the error between the estimated values and our imputed values. Note our imputed values refer to $y$ values and are taken as actual labels (values) to the machine learning model while the estimated values refer to $\hat{y}$ values obtained by the XGBoost model. Thus, the estimated values are different that our imputed values derived from our methods presented in previous sections.

Using the proper definitions from~\cite{xgboost-Chen-Guestin}, let $\D = \{(X_i,y_i)\}$ be a dataset with $n$ samples each of $d$ features, where $|\D| = n,~X_i \in \reals^d,~y_i \in \reals$. A tree ensemble model is comprised of $K$ additive predictive functions as:

$$\hat{y}_i = \theta(X_i) = \sum_{k=1}^K f_k(X_i),~~f_k \in \F,$$ 

where $\F$ is the
space of classification and regression trees, and $\theta$ is the set of parameters. In the training, the model amounts to finding the best parameters $\theta$ that best fit the training data $X_i$ and $y_i$ labels. Here $X_i$ is the feature set as described in Section~\ref{sec:data} and $y_i$ is the target, i.e., delivery frequency that the model aims to estimate for the $i$th sample ($s_i$) in $\D$.
As introduced in~\cite{xgboost-Chen-Guestin} a salient characteristic of objective function is that it consists of two parts: training loss and regularization term:

$$L(\theta) = \sum_{i=1}^nl(y_i, \hat{y}_i) + \sum_{k=1}^K \Omega(f_k).$$

Here $l$ is a loss function that measures
the difference between the prediction $\hat{y}_i$ and the target $y_i$ and $\Omega$ penalizes the complexity of the model. Using the above function we learn appropriate $\theta$ for our data and we reveal the features that most significantly contribute to our prediction. Further details on gradient descent algorithm is brought in~\cite{xgboost-Chen-Guestin}.

This model assists us with explaining our data and illuminating on covariates affecting our estimations the most. In particular, we use Shapley values in XGBoost; a Shapley value is the average expected marginal contribution of one covariate after all possible combinations have been considered by the model. For further details on Shapley values see~\cite{shapley}.
\begin{figure} [htpb]
     \centering
     \begin{subfigure}[htpb]{0.49\textwidth}
         \centering
         \includegraphics[width=\textwidth]{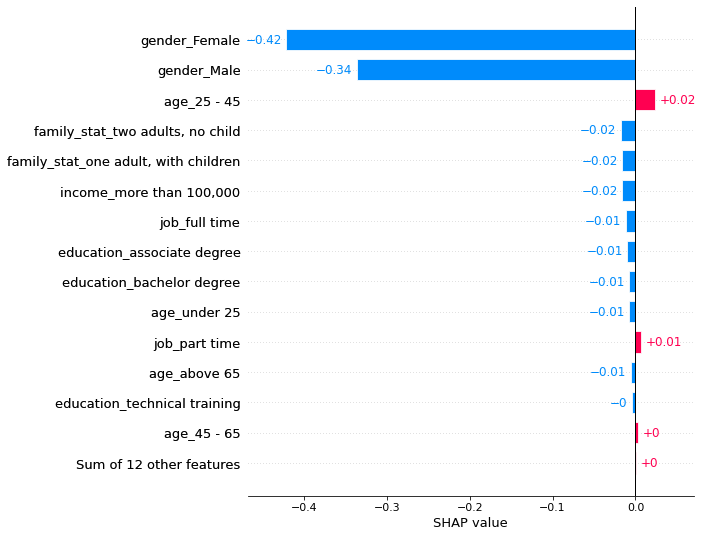}
         \caption{PSRC 2017}
         \label{fig:expl1}
     \end{subfigure}
     \hfill
     \begin{subfigure}[htpb]{0.49\textwidth}
         \centering
         \includegraphics[width=\textwidth]{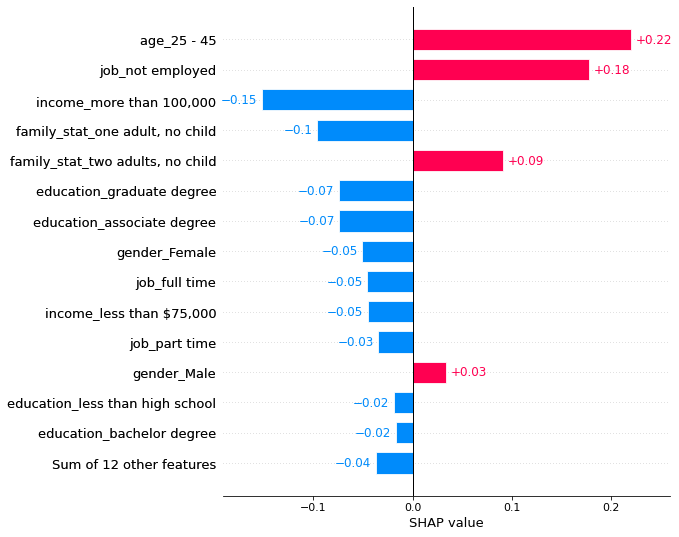}
         \caption{NHTS 2017}
         \label{fig:expl2}
     \end{subfigure}
     \hfill
     \begin{subfigure}[htpb]{0.49\textwidth}
         \centering
         \includegraphics[width=\textwidth]{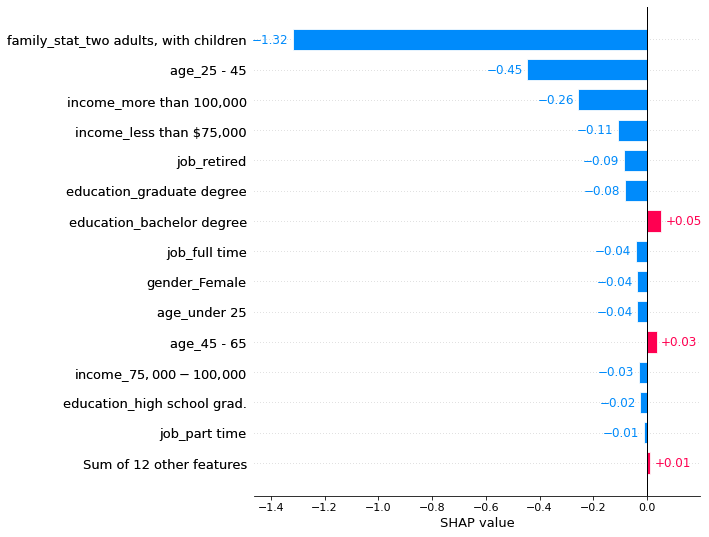}
         \caption{PSRC 2021}
         \label{fig:expl3}
    \end{subfigure}
    \hfill
     \begin{subfigure}[htpb]{0.48\textwidth}
         \centering
         \includegraphics[width=\textwidth]{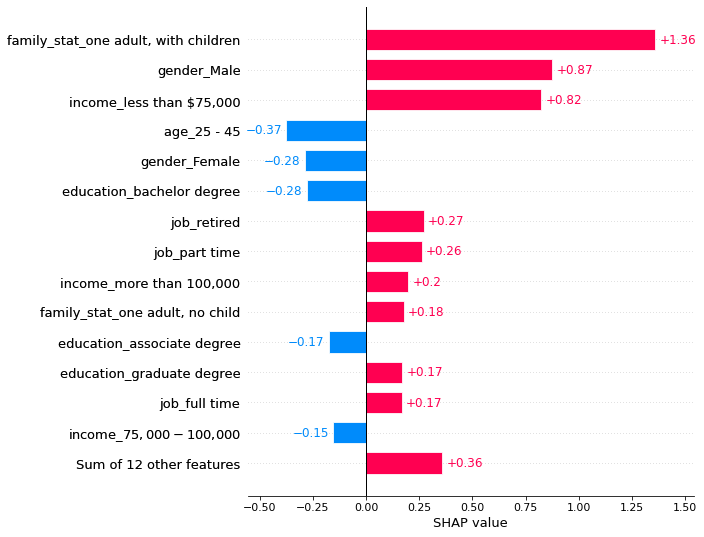}
         \caption{NHTS 2021}
         \label{fig:expl4}
     \end{subfigure}
        \caption{Covariates that contribute the most in estimations in all four datasets. Those features (and categories) with marginal contribution highlighted in red contribute to more delivery volumes and the ones in blue constitute less delivery volumes.}
        \label{fig:exploratory}
\end{figure}
We systematically run our imputed datasets; PSRC 2017 and sythetic NHTS-cut 2021 together with NHTS-cut 2017 and PSRC 2021 through the XGBoost model and we derive the Shapley values of the features in each dataset exclusively.

As we can observe in Figure~\ref{fig:exploratory}, our generated synthetic dataset (NHTS-cut 2021) inherits marginal contributions from all three datasets due to the Nested Nearest Neighbor Matching employed in Algorithm~\ref{alg:generate}. In particular, those households of the income $\$75,000 - \$100,000$ and individuals between the age of 25 -- 45 in both 2021 datasets is contributing to deliveries of fewer volumes. Such individuals in 2017 datasets, however, are contributing to deliveries with higher amounts. 
Conversely, those households with income higher than $\$100,000$ are expected to have more deliveries as demonstrated in synthetic NHTS-cut 2021. In NHTS 2017 and 2021 both male individuals are the most consumers of at home deliveries, while this is the opposite in PSRC 2017 and 2021. In terms of family life-cycle, households with one adult have more at-home delivery in our synthetic dataset, which is inconsistent with with the marginal contribution of one-adult households in all three datasets. In terms of employment, most job categories in the synthetic dataset contribute to higher delivery volumes, while only part-time and job-not employed categories had a positive impact on deliveries in PSRC 2017 and NHTS 2017, respectively. 

Overall, we conclude that our synthetic dataset attained the spirit of the data from all three dataset conceptually, which can be considered a reasonable first step toward forecasting future data values given limited historical data in the past, especially considering the imputation and harmonization steps taken to achieve this result.

\section{Conclusion} \label{sec:conclusion}

In this paper we developed frameworks and quantitative methods for estimating household delivery demand, and to further understand the important parameters influencing the number of deliveries per household. We addressed the issue of data scarcity by matching two publicly available data sources– the 2017 Puget Sound Regional Council (PSRC) Travel Survey, and the 2017 National Household Travel Survey (NHTS). While both contained our dependent variable of interest, number of deliveries per household, the majority of delivery data were missing in the PSRC 2017 dataset and complete in NHTS 2017.  Making use of a near-ground-truth data source (NHTS 2017), we propose an imputation method to obtain the missing values in PSRC 2017. Moreover, our imputation method provides a similarity matching between the two data sources’ samples. Further, we used the imputation by matching approach to generate a synthetic NHTS 2021 dataset. This approach affords us much flexibility in addressing the data challenges and provides more accuracy with our online purchase demand estimation. Exploratory analysis of the results using a predictive algorithm shows a high predictive accuracy between the ground-truth and synthetic dataset as some of the marginal contributions of the synthetic dataset inherited the marginal contribution of the features of those samples in the three datasets due to the nested nearest matching we used. The results were evaluated and found to be realistic.  

This method offers a novel approach to the problem of forecasting demand for household delivery in small areas based upon publicly available data (e.g. zip code). It has important applications for assessing business opportunities, for example home delivery operations, as well as public policy and planning, such as traffic forecasting. However, since this method includes imputation and harmonization procedures, results should be interpreted with caution. Imputation can introduce systematic sources of bias, for example, survey response bias would tend to overestimate deliveries for households that participated in the household travel survey and underestimate deliveries for households that did not.  


%






%

\bibliography{trb_template}
\bibliographystyle{unsrt}




%








\end{document}